\documentclass[12pt]{iopart}
\usepackage{graphicx}
\usepackage{cite}
\usepackage{amsfonts}
\usepackage{wrapfig}
\usepackage{srcltx}
\usepackage{color}

\begin{document}
\title[Mutual Information as a Two-Point Correlation Function]
      {Mutual Information as a Two-Point Correlation Function in Stochastic Lattice Models}
\author{Ulrich M\"uller and Haye Hinrichsen}
\address{Universit\"at W\"urzburg, Fakult\"at f\"ur  Physik und Astronomie, 97074  W\"urzburg, Germany.}

\begin{abstract}
In statistical physics entropy is usually introduced as a global quantity which expresses the amount of information that would be needed to specify the microscopic configuration of a system. However, for lattice models with infinitely many possible configurations per lattice site it is also meaningful to introduce entropy as a local observable that describes the information content of a single lattice site. Likewise, the mutual information can be interpreted as a two-point correlation function. Studying a particular growth model we demonstrate that the mutual information exhibits scaling properties that are consistent with the established phenomenological scaling picture.
\end{abstract}

\def\d{{\rm d}}
\def\0{\emptyset}

\ead{ulrich.mueller@physik.uni-wuerzburg.de}
\ead{hinrichsen@physik.uni-wuerzburg.de}

\parskip 2mm 

\section{Introduction}

In statistical mechanics the physical properties of fluctuating systems are usually described in terms of correlation functions which are defined as the expectation values of products of observables located at different points in space and/or time. For example, the critical equilibrium state of the Ising model is known to be characterized by spin-spin correlations of the form
\begin{equation}
G(i,j)=\langle s_is_j \rangle\sim |i-j|^{2-d-\eta}.
\end{equation}
Here $s_i=\pm 1$ is the local classical Ising spin at site $i$, $\langle\ldots\rangle$ denotes the ensemble average, $d$ is the dimension, and $\eta$ is the associated critical exponent. 

Another important pillar of statistical physics is the concept of entropy which describes the information content of a system. More specifically, if a randomly evolving system is characterized by a set of possible configurations (microstates) $c \in \Omega$ with a probability distribution $P(c)$, the amount of information needed to specify a particular configuration $c$ (in bit times $\ln 2$) is given by 
\begin{equation}
S(c)=-\ln P(c) \,.
\end{equation}
Averaging over all configurations one obtains the Boltzmann-Gibbs or Shannon entropy
\begin{equation}
S=\langle S \rangle = -\sum_{c} P(c) \ln P(c)\,,
\end{equation}
which describes the mean information content of the system. For example, in the equilibrium state of the Ising model, where the probability of a configuration $c=\{s_i\}$ is given by the Boltzmann weight $P(c) = Z^{-1}e^{-\beta H(c)}$ normalized by the partition sum $Z$, the average entropy is given by $\langle S \rangle = \ln Z + \beta \langle E \rangle$. Clearly, entropy as defined above is a quantity that characterizes the system globally.

In the present work we suggest to look at entropy from a different perspective: Instead of defining entropy as a \textit{global} quantity $S$, we want to use it as a \textit{local} observable~$S_i$ which describes the information content  of a microscopic portion of the system, in the simplest case the information of a single site $i$. Such a local entropy is particularly interesting in models with infinitely many possible configurations per lattice site.\footnote{If there were only a finite number of possible configurations per site the local entropy would be just a linear combination of ordinary local observables which is not expected to yield new insights. For example, in the case of Ising spins, where each site carries no more than one bit, the local entropy could take only two values and hence can be expressed in terms of the spin variables.} As a natural candidate, we will study here a particular growth process, where the height above a lattice site is unrestricted. 

The local entropy $S_i$, as will be defined below, can be viewed as a special kind of one-point function. Likewise it is possible to study the joint entropy $S_{ij}$ at two different lattice sites $i$ and $j$. If these sites are uncorrelated one expects that $S_{ij}=S_i+S_j$. Therefore, it is useful to consider the ``connected part''
\begin{equation}
\label{mutual}
I(i,j) = S_i + S_j - S_{ij}
\end{equation}
which is known as the \textit{mutual information} in information theory~\cite{MacKay}. Roughly speaking $I(i,j)$ quantifies how much information site $i$ has about the state of site $j$ and vice versa. This concept can be easily generalized to $n$-point functions by considering the corresponding multivariate mutual information. Note that this concept differs from previous studies, where the mutual information between sections of a bipartite system was studied~\cite{Wilms}.

Entropic correlation functions like the mutual information differ from ordinary correlation functions insofar as the logarithm is a nonlinear function and therefore involves arbitrary high powers of the local field variables. For this reason it is not obvious whether such a correlation function exhibits the same type of phenomenological scaling laws as ordinary ones in the vicinity of a phase transition. However, the results of the present work suggest that it is possible to establish a set of consistent scaling laws.

\section{Definition and properties of the growth process}

\subsection{Definition}

\begin{figure}
\centering\includegraphics[width=120mm]{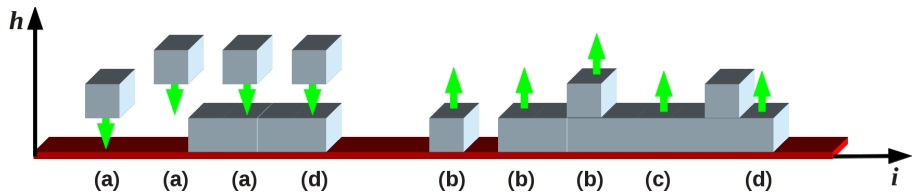}
\caption{Dynamical rules of the growth process. (a) Particles are deposited everywhere at rate $q$. (b) Solitary particles and particles at the edges of plateaus evaporate at rate $1$. (c) Particles from the middle of a plateau desorb at rate $p$. (d) Deposition and evaporation is forbidden if the resulting configuration would violate the RSOS constraint (\ref{RSOS}).}
\label{fig:model}
\end{figure}

As an example of a model with infinitely many possible configurations per lattice site, we study a simple solid-on-solid growth process which was discussed some time ago in the context of non-equilibrium wetting~\cite{wetting1,wetting2}. The model is defined on a one-dimensional periodic lattice with $L$ sites labeled by $i=1,\ldots,L$. Each site carries an unbounded variable $h_i=0,1,\ldots$ which describes the height of an interface above an inert substrate. Moreover, an effective interaction (surface tension) is introduced by imposing the so-called restricted solid-on-solid (RSOS) condition
\begin{equation}
\label{RSOS}
|h_i - h_{i\pm 1}| \leq 1\,,
\end{equation}
i.e. neighboring sites must not differ by more than one unit. The model evolves by random-sequential deposition and evaporation constrained by this condition (see Fig.~\ref{fig:model}). It is controlled by two parameters, namely, a growth rate $q$, and another parameter $p$ for the desorption from the middle of plateaus which allows one to interpolate between equilibrium and non-equilibrium (see Ref. \cite{wetting2} for further details).

\subsection{Scaling properties}

The growth model defined above is known to exhibit a continuous phase transition from a bound to a moving phase at a particular threshold $q=q_c(p)$. This transition can be described in terms of two different order parameters, namely, the interface width $w$ defined as the standard deviation of the height, and the density $n_0$ of contact points at the bottom layer. Moreover, the critical behavior is characterized by a typical correlation length $\xi_\perp$ and correlation time $\xi_\parallel$. In the stationary bound phase close to the transition, where the critical parameter
\begin{equation}
\label{critpar}
\epsilon = q_c(p) -q
\end{equation}
is small and positive, these quantities scale as
\begin{equation}
\label{StatScaling}
w \sim \epsilon^{-\zeta}\,,\qquad n_0 \sim \epsilon^\beta\,,\qquad  \xi_\perp \sim \epsilon^{-\nu_\perp}\,,\qquad  \xi_\parallel \sim \epsilon^{-\nu_\parallel}\,,
\end{equation}
provided that the system size $L \gg \xi_\perp$ is large enough. At the critical point $q=q_c$ ($\epsilon=0$) one finds instead an asymptotic time dependence of the form
\begin{equation}
\label{TimeScaling}
w \sim t^{\alpha/z} \,,\qquad n_0 \sim t^{-\theta} \,,\qquad \xi_\perp \sim t^{1/z}\,, \qquad
\xi_\parallel \sim t\,,
\end{equation}
where $\alpha=\zeta/\nu_\perp$, $\theta=\beta/\nu_\parallel$, and $1/z = \nu_\perp/\nu_\parallel$. Starting with a flat initial state in the bound phase near the critical point one observes a crossover from (\ref{TimeScaling}) to (\ref{StatScaling}) which can be expressed by scaling forms with certain universal scaling functions. For example, one finds that the interface width grows with time according to the scaling form
\begin{equation}
w(t,\epsilon) \;\simeq\; t^\alpha \, W(t \,\epsilon^{\nu_\parallel})\,.
\end{equation}
The values of the critical exponents and the scaling functions are determined by the universality class of the phase transition. In the present model the parameter $p$ allows one to choose between three different classes, namely, the bounded KPZ class with positive and negative nonlinearity (bKPZ$\pm$) for $p>1$ and $p<1$, as well as the bounded Edwards Wilkinson class (bEW) for $p=1$. The expected values of the critical exponents and the corresponding critical thresholds are listed in Table \ref{tab:exponents}.

\begin{table}[t]
\centering
\vspace{+0.1cm}
\begin{tabular}{|c||c|c||c|c|c|c|c|c|c|}
\hline
class 	 & $p$ & $q_c(p)$    &$\alpha$&$z$  &$\nu_\perp$&$\nu_\parallel$&$\zeta$&$\theta$   & $\beta$ \\
\hline
bKPZ+  & 0.1 &    0.61117(1)  &$1/2$   &$3/2$&$1$ 	  &$3/2$  	  &$1/2$  &$1.184(10)$& $1.776(15)$\\
bEW    & 1   &    1           &$1/2$   &$2$  &$2/3$      &$4/3$  	  &$1/3$  &$3/4$      & $1$\\
bKPZ-  & 2.0 &    1.23237(1)  &$1/2$   &$3/2$&$1$        &$3/2$  	  &$1/2$  &$0.228(5)$ & $0.342(8)$\\
\hline
\end{tabular} 
\caption{Values of the critical parameters and the expected critical exponents~\cite{barato} of the growth process in 1+1 dimensions. The exponents are related by the scaling relations $\theta=\beta/\nu_\parallel$, $\alpha=\zeta/\nu_\perp$, $z=\nu_\parallel/\nu_\perp$, and in 1+1 dimensions by $\nu_\parallel=\zeta+1$.}
\label{tab:exponents}
\end{table}

\subsection{Exact solution for $p=1$}

For $p=1$ and $q<1$ the model defined above is known to relax into an Boltzmann-distributed equilibrium state obeying detailed balance. This state is characterized by the partition sum
\begin{equation}
\label{Z}
Z \;=\; \sum_{h_1}\ldots\sum_{h_L}  \, \prod_{i=1}^L q^{h_i} \;=\; \sum_{\{h\}} q^{\sum_{i=1}^L h_i}
\end{equation}
which runs over all configurations compatible with the RSOS constraint (\ref{RSOS}). To see this note that in a model \textit{without} the RSOS constraint (\ref{RSOS}) each site would independently perform a bounded biased random walk in height direction. These decoupled random walks would evolve into a stationary state obeying detailed balance, where the probability of finding the value $h$ is proportional to~$q^h$. Clearly, such a decoupled system would be described by the partition sum~(\ref{Z}) with unrestricted summation. Then, imposing the additional constraint (\ref{RSOS}), it is easy to see that detailed balance is not violated and that the Boltzmann weights are preserved, -- the only thing what changes is the summation (\ref{Z}) which is now restricted to configurations satisfying the constraint (\ref{RSOS}).

The stationary state for $p=1$ can be described in terms of a transfer matrix formalism~\cite{transfermat1,transfermat2,wetting2} by reorganizing the partition sum (\ref{Z}) as
\begin{eqnarray}
Z &=& \sum_{h_1}\ldots\sum_{h_L} \prod_{i=1}^L q^{(h_i+h_{i+1})/2} \\
&=& \sum_{h_1} q^{(h_1+h_2)/2} \sum_{h_2} q^{(h_2+h_3)/2}\ldots \sum_{h_L} q^{(h_L+h_1)/2} \nonumber \\
&=& \sum_{h_1} T_{h_1,h_2} \sum_{h_2} T_{h_2,h_3} \ldots \sum_{h_1} T_{h_L,h_1} \;=\; \Tr[T^L]   \,, \nonumber \\
\end{eqnarray}
where $T$ is the transfer matrix with the infinite-dimensional tridiagonal representation
\begin{equation}
T = \left(
\begin{array}{cccccc}
1 & q^{1/2} &&&& \\
q^{1/2} & q & q^{3/2} && \\
& q^{3/2} & q^2 & q^{5/2} && \\
&& ... & ... & ... & \\
\end{array}
\right)\,.
\end{equation}
Defining canonical basis vectors $|h\rangle$ and $\langle h |$ the probability of finding site $i$ at height $h$ is then given by
\begin{equation}
\label{expval}
P(h) \;=\; \frac{\Tr\biggl[ T^{i} |h\rangle\langle h| T^{L-i}\biggr] } {\Tr[T^L]} \;=\; \frac{\langle h|T^L| h \rangle}{Z}\,.
\end{equation}
The transfer matrix $T$ is symmetric and has a non-degenerate spectral decomposition
\begin{equation}
\label{specdec}
T \;=\; \sum_{n=0}^\infty \lambda_q^{(n)} |\phi_q^{(n)}\rangle\langle\phi_q^{(n)}|
\end{equation}
with real eigenvalues $\lambda_q^{(n)}$ and pairwise orthonormal eigenvectors $|\phi_q^{(n)}\rangle$ and $\langle\phi_q^{(n)}|$. Since a high power of such a matrix is dominated by its largest eigenvalue, we may therefore approximate $T^L$ in the thermodynamic limit $L \to \infty$ by
\begin{equation}
T^L \approx \lambda_q^L \, |\phi_q\rangle\langle \phi_q|\,,
\end{equation}
where $\lambda_q \equiv \lambda_q^{(0)}$ denotes the largest eigenvalue of $T$ with the corresponding eigenvectors  $\langle \phi_q |$ and  $|\phi_q\rangle$. Consequently $Z\approx\lambda_q^L$ so that the expectation value of finding a site at height $h$ is given by
\begin{equation}
P(h) \;=\;  \langle h | \phi_q \rangle \langle \phi_q | h \rangle \;=\; |\langle h | \phi_q \rangle|^2\,.
\end{equation}
Remarkably, the transfer formalism reminds one of the Dirac formalism in quantum mechanics although the present problem is classical.

For general $q<1$ the determination of the dominating eigenvector is non-trivial and to our knowledge a closed solution is not yet known. However, close to the transition, where $\epsilon=1-q$ is small, the eigenvector $\phi_q(h)$ can be approximated by an Airy function of the form~\cite{wetting2}
\begin{equation}
\label{Airy}
\phi(h)\,\simeq\,\frac{(3\epsilon)^{1/6}}{\sqrt{A}} \, \mbox{Ai}\Bigl[ (3\epsilon)^{1/3}h + z_0\Bigr]\,,
\end{equation}
where $z_0 \approx -2.33811$ is the largest root of Ai$(z)$ and $A =\int_{z0}^\infty \mbox{Ai}^2(z) \d z \approx 0.491697$ is the corresponding normalization.

\section{One-point function: Local entropy}

In a growth process the local entropy at site $i$ is given by
\begin{equation}
\label{OnePoint}
S_i \;=\; -\sum_{h_i} P(h_i) \ln P(h_i)\,.
\end{equation}
Near criticality, where $\epsilon=q_c-q$ is small, the probability $P(h_i)=P(h)$ to find the interface at height $h$ is expected to obey the scaling form
\begin{equation}
\label{ScalingProfile}
P(h) \;\simeq\; \epsilon^\zeta\, \Phi(h \epsilon^\zeta)\,,
\end{equation}
where $\Phi(z)$ is a scaling function determined by the universality class selected by $p$. Replacing the sum in (\ref{OnePoint}) by an integral and inserting this scaling form one can show that the one-point entropy scales as
\begin{equation}
\label{I0}
S_i \simeq C - \zeta \ln \epsilon
\end{equation}
in the limit $\epsilon\to 0$, where
\begin{equation}
C = - \int_{0}^\infty \, \Phi(z) \ln \Phi(z) \d z
\end{equation}
Since the universal scaling function $\Phi(z)$ is only defined up to a rescaling the constant $C$ is non-universal. For $p=1$, where the exact solution (\ref{Airy}) leads to the scaling function $\Phi(z) = 3^{1/3} A^{-1} \mbox{Ai}^2(3^{1/3}z+z_0)$, we obtain the numerical value $C\simeq 0.650832$. 

\begin{figure}
\centering\includegraphics[width=100mm]{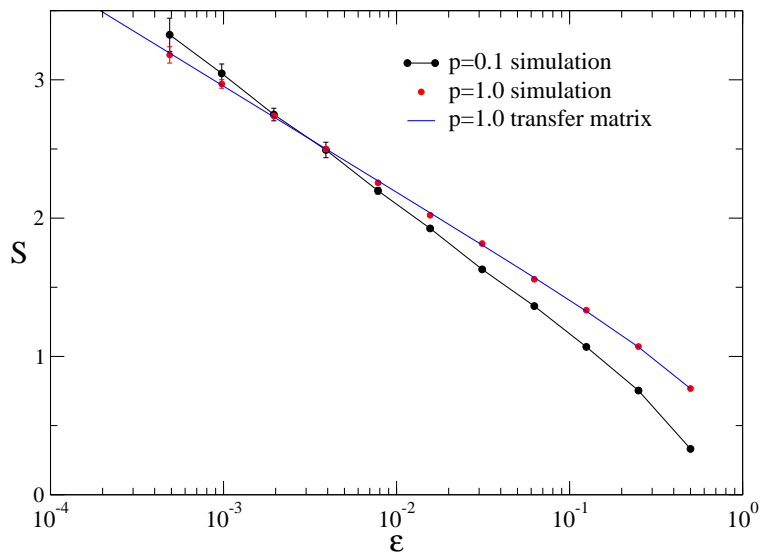}
\caption{Local entropy as a one-point function in the stationary state (see text).}
\label{fig:entropy}
\end{figure}

To confirm the predicted scaling behavior, we measured the local entropy in a numerical simulation (see Fig.~\ref{fig:entropy}). In the Edwards-Wilkinson case $p=1$ the numerical data (red dots) agree very well with the transfer matrix results. Moreover, the logarithmic decay with a slope $-0.34$ is in agreement with the expected exponent $\zeta=1/3$. For $p=0.1$ the measured slope $-0.43$ is not in full agreement with the expected exponent $\zeta=1/2$ of the bKPZ- class. This confirms that the crossover from EW to KPZ behavior of this particular model is very slow, see Ref.~\cite{barato} for a detailed discussion.

\section{Two-point function: Mutual information}

By means of the RSOS constraint (\ref{RSOS}) the lattice sites are not independent ,but exchange some information about their local state. As outlined in the introduction, this information exchange is most naturally quantified by the mutual information $I(i,j) = S_i + S_j - S_{ij}$ between two sites $i$ and $j$. Because of periodic boundary conditions the mutual information will only depend on the distance $r=|i-j|$ between the points, i.e., $I(i,j)=I(r)$. 

In the special case of $p=1$ the transfer matrix formalism provides a tool to calculate the mutual information analytically. To this end one has to compute the joint entropy
\begin{equation}
S_{ij} \;=\; -\sum_{h_i,h_j} P(h_i,h_j) \ln P(h_i,h_j)
\end{equation}
in terms of the joint probability $P(h_i,h_j)$. This probability is given by 
\begin{equation}
\label{pij}
P(h_i,h_j)= Z^{-1} \sum_{\{h\}_{ij}} q^{\sum_{k=1}^L h_k}
\end{equation}
were the sum runs over all possible configurations while keeping the heights $h_i $ and $h_j$ at the positions $i$ and $j$ fixed. Using the transfer matrix method these probabilities can be expressed as
\begin{equation}
\label{pij}
P(h_i,h_j)= \frac{ \langle \phi_q | h_i \rangle  \langle h_i | T^{|i-j|}  | h_j \rangle \langle h_j  | \phi_q \rangle} {\lambda_q^{|i-j|}}.
\end{equation}
Although it is not trivial to calculate the mutual information from this expression, one can find useful approximations in the limit of short as well as very large distances. 

\noindent\textbf{Short distance limit:}\\
If the distance between the two points is much shorter than the correlation length, one can  estimate the decay of the mutual information as follows. First note that the mutual information can be written as
\begin{equation}
\label{mi_m1}
I(i,j)\;=\; S_i - S_{i|j}
\end{equation}
with the conditional entropy

\begin{equation}
\label{conent}
S_{i|j} \;=\; -\sum_{h_i} P(h_i) \sum_{h_j} P(h_j|h_i) \ln P(h_j|h_i).
\end{equation}
For neighboring sites the conditional probability to find sites $i+1$ at height $h_{i+1}$ given that site $i$ is at height $h_i$ reads
\begin{equation}
\label{neighbour}
P(h_{i+1}|h_i) \;=\; \frac{P(h_i,h_{i+1})}{P(h_i)} \;=\; \left\{
\begin{array}{ll}\frac{q^{\frac{h_i+h_{i+1}}{2}}}{\lambda_q}\frac{\langle h_{i+1} | \phi_q \rangle}{\langle h_{i} | \phi_q \rangle} & \mbox{if $|h_i-h_{i+1}| \le 1$} \\ 
0 & \mbox{otherwise.}
\end{array}
\right.
\end{equation}
In the limit $q \to 1$, where $\langle h_{i+1} | \phi_q \rangle \simeq \langle h_{i} | \phi_q \rangle$, this expression reduces to
\begin{equation}
P(h_{i+1}|h_i) \;=\; \left\{
\begin{array}{ll}1/3  & \mbox{if $|h_i-h_{i+1}| \le 1$} \\ 
0 & \mbox{otherwise.}
\end{array}
\right.
\end{equation}
This shows that on short distances the interface height $h_i$ increases or decreases by one unit or stays at the same height with equal probability as we move to the neighboring lattice site. In other words, on short distances the interface describes an unbiased random walk in height direction. Therefore, if the distance $r=|i-j|$ is sufficiently larger than 1, but still smaller than the correlation length, the central limit theorem implies that the conditional probability $P(h_j|h_i)$ is approximately given by a normal distribution centered around $h_i$ with the width proportional to $\sqrt{r}$. Consequently the conditional entropy is of the form
\begin{equation}
\label{normalentropy}
H(h_i|h_j) \;\simeq\; H_0 + \frac{1}{2}\ln(r)
\end{equation}
with the numerical offset $H_0=1.216206$. Inserted into~(\ref{mi_m1}) and using (\ref{I0}) this leads to
\begin{equation}
\label{shorteq}
I(r) \;\simeq\; I_0-\frac{1}{3}\ln(\epsilon)-\frac{1}{2}\ln(r)
\end{equation}
with the numerical value $I_0=-0.565374$. 

\noindent\textbf{Long distance limit:}\\
In the limit where $r=|i-j|$ is much larger than the correlation length, the two sites are almost statistically independent so that the joint probability distribution $P(h_i,h_j)$ differs only slightly from $P(h_i)P(h_j)$, i.e.
\begin{equation}
P(h_i,h_j) \;=\; P(h_i)P(h_j) + \eta_{h_i,h_j}\,,
\end{equation}
where $\eta_{h_i,h_j}\ll 1$. This allows the mutual information to be expanded as
\begin{eqnarray}
\label{expansion}
\fl\qquad 
I(i,j)=S(h_i) + S(h_j)- S(h_i, h_j) \nonumber \\[2mm]
\fl\qquad \qquad\quad  = -\sum_{h_i,h_j}P(h_i)P(h_j)\ln [P(h_i)P(h_j)] + \sum_{h_i,h_j}P(h_i,h_j)\ln P(h_i,h_j)\\
\fl\qquad \qquad \quad = \sum_{h_i,h_j} \Bigl[\biggl(1+\ln P(h_i)P(h_j)\biggr) \eta_{h_i,h_j} +
\frac{\eta^2_{h_i.h_j}}{2P(h_i)P(h_j)}\Bigr] + \mathcal{O}(\eta^3)\,.\nonumber
\end{eqnarray}
To compute the small deviation $\eta(h_i,h_j)$, we insert the spectral decomposition (\ref{specdec}) into Eq. (\ref{pij}) one obtains 
\begin{eqnarray} 
\fl\qquad
P(h_i,h_j) \;=\; \underbrace{\langle \phi_q|h_i\rangle^2 \langle \phi_q|h_j\rangle^2}_{=P(h_i)P(h_j)}+\sum_{n=1}^{\infty} \frac{\langle \phi_q|h_i\rangle\langle h_i|\phi_q^{(n)} \rangle (\lambda_q^{(n)})^{|i-j|} \langle \phi_q^{(n)} | h_j\rangle \langle h_j | \phi_q\rangle}{\lambda_q^{|i-j|}}\,.
\end{eqnarray}
Thus we can identify $\eta_{h_i,h_j}$ with the first summand, i.e.
\begin{equation}
\eta_{h_i,h_j} \;=\; \langle \phi_q|h_i\rangle\langle h_i|\phi_q^{(1)} \rangle  \langle \phi_q^{(1)} | h_j\rangle \langle h_j | \phi_q\rangle \, \Delta^{-|i-j|}
\end{equation}
where $\Delta=\lambda^{(0)}_q/\lambda^{(1)}_q$ is the gap ratio between the leading and the next-to-leading eigenvalue. Inserting this expression back into the expansion (\ref{expansion}) one can show by using the orthogonality of the eigenvectors $\langle \phi_q|\phi_q^{(1)}\rangle=0$ that the first-order contribution vanishes. Therefore, to second order in $\eta_{h_i,h_j}$ the mutual information is given by
\begin{equation}
I(r) \;\simeq\;\frac12 \underbrace{\sum_{h_i,h_h} \langle h_i|\phi_q^{(1)} \rangle^2  \langle \phi_q^{(1)} | h_j\rangle^2}_{=1} \,  \Delta^{-2r}
\;=\; \frac{1}{2}e^{-2r\ln \Delta}\,.
\end{equation}
This means that in the long-distance limit the mutual information decays exponentially as $I(r) \sim e^{-r/\xi_\perp}$ with the correlation length
\begin{equation}
\label{longeq}
\xi_\perp = \frac{1}{2 \ln \Delta}\,.
\end{equation}
The correlation length, which is determined by the first gap ratio $\Delta$ of the transfer matrix, depends on $\epsilon=q_c-q$. As expected, one finds numerically that $\xi_\perp \sim \epsilon^{-\nu_\perp}$ in agreement with the scaling behavior of Eq.~(\ref{Airy}).

\newpage
\noindent\textbf{Scaling form:}\\
The two asymptotic formulas (\ref{shorteq}) and (\ref{longeq}), which both depend on the scale-invariant ratio $r/\xi_\perp$, suggest that the crossover from one behavior to the other is given by a scaling law of the form
\begin{equation}
\label{scalingform}
I(r) \;\simeq\; F(r/\xi_\perp) \,,
\end{equation}
where $F$ is a scaling function which is expected to be universal. Because of (\ref{shorteq}) and (\ref{longeq}) this scaling function behaves asymptotically as
\begin{equation}
F(z) \;\simeq\; \left\{
\begin{array}{ll}
I_0 - \frac12 \ln z & \mbox{ for } 1 \ll r \ll \xi_\perp \\
\frac12 \exp(-z/A) & \mbox{ for } \qquad r \gg \xi_\perp
\end{array}
\right.\,.
\end{equation}
Note that in contrast to conventional scaling forms, there is no leading power law in front of the scaling function $F$ in Eq. (\ref{scalingform}).

\begin{figure}
\centering\includegraphics[width=140mm]{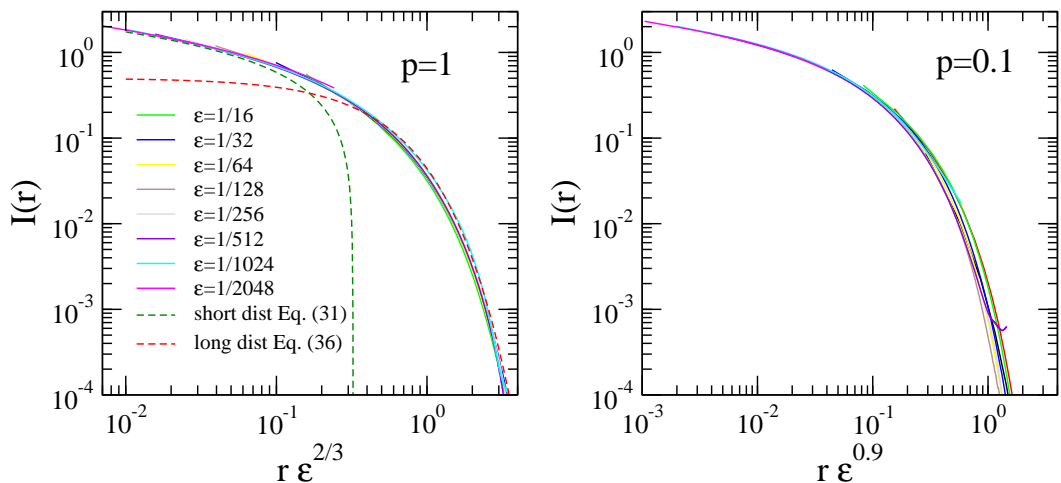}
\caption{Mutual information $I(r)$ plotted as a function of the scale-invariant combination $r e^{\nu_\perp}$, leading to a data collapse according to the scaling form (\ref{scalingform}). Left: Numerical data for the bounded Edward-Wilkinson case $p=1$ together with the short- and long-time approximations in Eqs.~(\ref{shorteq}) and (\ref{longeq}) shown as dashed lines. Right: Corresponding data collapse in the bKPZ case $p=0.1$ (see text). }
\label{fig:mutual}
\end{figure}

To test this hypothesis we measured the mutual information as a function of $r$ in a numerical simulation for various values of $\epsilon$. As shown Fig.~\ref{fig:mutual} one obtains a convincing data collapse in the case $p=1$. Similar results are obtained by using the transfer matrix formalism. For $p=0.1$ the best possible data collapse is obtained for $\nu_\perp \approx 0.9$ which differs from the expected value $\nu_\perp=1$ for the bKPZ- class. This discrepancy is again caused by the slow crossover from EW to KPZ in this model.

\section{Conclusions}

In stochastic lattice models the local entropy describes the information content or uncertainty of the local state of a single lattice site. Likewise the mutual information describes how strongly two sites are correlated. These quantities are particularly interesting in models with infinitely many states per site, where they cannot expressed as finite linear combinations of ordinary correlation functions. This leads to the question how entropic observables behave in systems with a continuous phase transition.

As an example, we have studied a simple growth model of a one-dimensional interface. In this model the interface at site $i$ is described by a local height $h_i=0,1,2,\ldots$ and thus it has infinitely many states per site. Moreover, the model exhibits an unbinding transition from the substrate controlled by the growth rate. Another parameter allows one to select various universality classes with different critical exponents.

As for the local entropy, interpreted here as a one-point function, we find a logarithmic scaling behavior of the form
\begin{equation}
S_i \simeq C - \zeta \ln \epsilon\,,
\end{equation}
where $\epsilon=q_c-q$ parametrizes the distance from criticality and $\zeta$ is one of the critical exponents listed in Table \ref{tab:exponents}. This result is expected since this exponent characterizes the width of the interface close to the transition.

The scaling behavior of the mutual information between two lattice sites, interpreted here as a two-point function, depends on the distance $r$ between the two points. For $p=1$ we find the asymptotic behaviors
\begin{equation}
\label{concmutual}
I(r) \simeq \left\{
\begin{array}{ll}
I_0-\frac12\ln r \epsilon^{\nu_\perp} & \mbox{ if } 1 \ll r \ll \xi_\perp \\
\frac12 e^{-r/\xi_\perp} & \mbox{ if } \qquad r \gg \xi_\perp
\end{array}
\right.
\end{equation}
where $\xi_\perp\sim\epsilon^{-\nu_\perp}$ denotes the correlation length. We expect these limits to remain valid in the KPZ case $p\neq 1$, using the corresponding KPZ exponents.

This asymptotic limits in Eq.~(\ref{concmutual}) suggest the general scaling form
\begin{equation}
I(r) \;=\; F(r \epsilon^{\nu_\perp})\,.
\end{equation}
In the present model this scaling form can be confirmed numerically, leading us to the conjecture that the scaling function $F$ is universal in the same sense as for ordinary correlation functions. However, in contrast to ordinary scaling functions, which usually describe the crossover between different power laws or the crossover from a power law to an exponential decay towards a constant, the function $F$ describes a crossover from a logarithmic to an exponential decay. 

Moreover, it is important to note that there is no leading power law in front of~$F$, meaning that the mutual information does not carry an intrinsic scaling dimension. Ordinary correlation functions carry an intrinsic scaling dimension which is usually determined by the scaling dimensions of the local observables. In entropic correlation functions, however, the logarithm involves arbitrary powers of local observables and therefore it is plausible that it cannot carry an intrinsic dimension. Whether or not this is a general feature of entropic correlation functions remains to be seen.

The proposed concept of entropic one- and two-point functions can easily be generalized to $n>2$ points by considering the so-called multivariate information between these points. Moreover, it is straight forward to apply similar ideas to quantum systems by replacing the local Shannon with the corresponding von-Neumann entropy.



\section*{References}
\begin{thebibliography}{99}

\bibitem{MacKay}
MacKay DJC, \textit{Information Theory, Inference, and Learning Algorithms}, Cambridge University Press, Cambridge, U.K. (2003).

\bibitem{Wilms}
Wilms J, Troyer M, and Verstraete F, \textit{Mutual information in classical spin models}, 2011 J. Stat. Mech: Theor. Exp. P10011.

\bibitem{wetting1}
Hinrichsen H, Livi R, Mukamel D, and Politi A, \textit{A Model for Nonequilibrium Wetting Transitions in Two Dimensions}, 1997 Phys. Rev. Lett. {\bf 79}, 2710. 

\bibitem{wetting2}
Hinrichsen H, Livi R, Mukamel D, and Politi A, \textit{Wetting under non-equilibrium conditions}, 2003 Phys. Rev. E \textbf{68}, 041606.

\bibitem{barato}
Barato AC, Hinrichsen H, and de Oliveira MJ, \textit{Numerical study of a model for nonequilibrium wetting},  2008 Phys. Rev. E \textbf{77}, 011101.

\bibitem{transfermat1}
van Leeuwen JMJ and Hilhorst HJ, \textit{Pinning of a rough interface by an external potential}, 1981 Physica A \textbf{107}, 319.

\bibitem{transfermat2}
Burkhardt TW,  \textit{Localisation-delocalisation transition in a solid-on-solid model with a pinning potential}, 1981 J. Phys. A: Math. Gen. \textbf{14}, L63. 

\end{thebibliography}
\end{document}